\documentclass[12pt]{article}

\normalsize

\let\oldtheequation=\theequation
\def\doteqs#1{\setcounter{equation}{0}
            \def\theequation{{#1}.\oldtheequation}}

\newcounter{sxn}
\def\sx#1{\addtocounter{sxn}{1}
\medskip \goodbreak
\noindent{\large\bf
\centerline{\thesxn.~~#1}} \nobreak \medskip}
\def\sxn#1{\sx{#1} \doteqs{\thesxn}}

\newcounter{axn}

\def\br{\begin{thebibliography}{abc}}
\def\er{\end{thebibliography}}

\begin{document}
\begin{flushright}
{SU/4240-718}
\end{flushright}
\thispagestyle{empty}
\centerline{\large\bf Noncommutative Geometry As A Regulator }
\bigskip
\begin{center}
Badis Ydri\footnote{E-mail: ydri@suhep.phy.syr.edu}.\\
\bigskip
{\it Physics Department , Syracuse University. }\\
\bigskip
{\it Syracuse , N.Y , 13244-1130 , U.S.A.}\\
\end{center}

\vskip.5cm
\begin{abstract}
We give a perturbative quantization of space-time $R^4$ in the
case where the commutators $C^{{\mu}{\nu}}=[X^{\mu},X^{\nu}]$ of
the underlying algebra generators are not central . We argue that
this kind of quantum space-times can be used as regulators for
quantum field theories . In particular we show in the case of the
${\phi}^4$ theory that by choosing appropriately the commutators
$C^{{\mu}{\nu}}$ we can remove all the infinities by reproducing
all the counter terms . In other words the renormalized
action on $R^4$ plus the counter terms can be rewritten as only a
renormalized action on the quantum space-time $QR^4$ . We
conjecture therefore that renormalization of quantum field theory
is equivalent to the quantization of the underlying space-time
$R^4$ .
\end{abstract}

\newpage
\sxn{Introduction}

Noncommutative geometry \cite{connes} allows one to define the
geometry of a given space in terms of its underlying algebra . It
is therefore more general than the ordinary differential geometry
in the sense that it enables us to describe algebraically the
geometry of any space whether or not it is  smooth and/or
differentiable . It is generally believed that NCG can be used to
reformulate if not to solve many problems in particle physics and
general relativity such as the problem of infinities in quantum
field theories and its possible connection to quantum gravity
\cite{coq,madore,landi,varilly,dop,kempf} . The potential of constructing
new nonperturbative methods for quantum field theories using NCG
is also well appreciated
\cite{coq,madore,landi,hcp,hcp1,hcp2,peter,bal,ydri,sachin,str,bars}
. The recent major interest in NCG however was mainly initiated by
the work of \cite{schwarz} on Yang-Mills theory on noncommutative
torus and its appearance as a limit of the matrix model of
M-theory . The relevance of NCG in string theory was further
discussed in \cite{witten} .

Quantum field theories on noncommutative space-time was
extensively analysed recently in the literature
\cite{seiberg,martin,krajewski,chep,ya,ben,mssw,hayakawa,adi,chms,cdp} and it
was shown that divegences although not completely removed they
are considerably softened . The reason is that the quantization
of $R^4$ or $R^2$ by replacing the coordinate functions $x^{\mu}$ by the
coordinate operators $X^{\mu}$ in the sense of \cite{dop} will
only modify vertices in the quantum theory and not propagators .
On compact spaces in the other hand such as the $4-$sphere $S^4$
\cite{hcp} , the $2-$ sphere $S^2$ \cite{hcp1} and $CP^2$
\cite{str} divegences are automatically cancelled out when we
quantize the space and that is because on comapct spaces (which was not the case on noncompact spaces) quantization leads to a finite number of degrees of freedom (points).

It is hoped that noncommutative geometry will shed new lights on
the meaning of renormalization because it provides a very
powerfull tools to formulate possible physical mechanisms
underlying the renormalization process of quantum field theories .
One such mecahnism which was developed by Deser \cite{deser} , Isham
et al.\cite{iss} and pursued in \cite{sorkin,madoregravity} is
Pauli's old idea that the quantization of gravity should give
rise to a discrete structure of space-time which will regulate
quantum field theories . As one can immediately see the typical
length scale of Pauli's lattice is of the order of Planck's scale
${\lambda}_p$ which is very small compared to the weak scale and
therefore corrections to the classical action will be very small
compared to the actual quantum corrections. This idea however is
still very plausible especially after the dicovery made in
\cite{seiberg} of an UV-IR mixing which could be used in a large
extra diemension-like activity to solve the above hierarchy
problem .

The philosphy of this paper will be quite different . We will
assume that space-time is really discrete and that the continuum
picture is only an approximation \cite{sorkin}. The discreteness however is
not given a priori but it is a consequence of the requirement that
the quantum field theory under consideration is finite . The
noncommutativity parameter ${\theta}$ is therefore expected to be
a function of both the space-time and the Quantum field theory
and it is completely determined by the finiteness requirement .
This simply mean that the quantization of space-time is achieved by
replacing the coordinate functions $x^{\mu}$ by the coordinate
operators $X^{\mu}$ as in \cite{snyder} but , and to the contrary to what was done in \cite{dop} , these operators will not satisfy the centrality
conditions $[X^{\mu},[X^{\nu},X^{\alpha}]]=0$ .

The paper is organized as follows : In section $1$ we introduce
the star product \cite{kon} for the case where the
noncommutativity parameter ${\theta}$ is not a constant . The
necessary and sufficient condition under which this star product
is associative turns out to be  simply
$[X^{\mu},[X^{\nu},X^{\alpha}]]=0$ . The associativity
requirement however is relaxed and allowed to be broken to the
first order in this double commutator . This relaxation is
necessary because one can check that we can not generalize
\cite{dop} , by making the commutators $[X^{\mu},X^{\nu}]$ not
central , while simultaneously preserving associativity . The
algebra $({\cal A},*)$ where ${\cal A}$ is the algebra of
functions on $R^4$ is then defined .

 In section $2$ we quantize
perturbatively the algebra $({\cal A},*)$ . In other words we
find the homomorphism $({\cal A},*)
{\longrightarrow}(A,{\times})$ order by order in perturbation
theory where $A$ is the algebra of operators generated by the
coordinate operators $X^{\mu}$ . The star product becomes under
quantization the nonassociative operator product ${\times}$ and
the corresponding Moyal bracket becomes the commutator
$[.,.]_{\times}$\cite{merkulov} . The difference between
${\times}$ and the ordinary dot product of operators is of the
order of the  double commutator$[X^{\mu},[X^{\nu},X^{\alpha}]]$ .
This is basically an example of deformation quantization
\cite{kon,merkulov,rieffel,gracia} and in particular it shows
explicitly the result of \cite{rieffel} that Doplicher et
al.\cite{dop} quantization prescription of space-time is a
deformation quantization of $R^4$  . We rederive also the
space-time uncertainty relations given in \cite{dop}  . In
section $3$ we construct a Dirac opertor on the quantum
space-time $QR^4$ , write down the action integrals of a scalar
field in terms of the algebra $(A,{\times})$ as well as in terms
of the algebra $({\cal A},*)$ . Finiteness requirement is then
used to fix ${\theta}$ in the two loops approximation of the
${\phi}^4$ theory  . Section $4$ contains conclusions and remarks
.

\sxn{The Star Product}

\vskip 5mm \noindent
{\bf 2.1 Associativity}
\vskip 5mm \noindent

Let $R^4$ be the space-time with the metric
${\eta}_{{\mu}{\nu}}=(1,1,1,1)$ . The algebra underlying the
whole differential geometry of $R^4$ is simply the associative
algebra ${\cal A}$ of functions $f$ on $R^4$ . It is generated by
the coordinate functions $x^{\mu}$ , ${\mu}=0,1,2,3$ . This
algebra is trivially a commutative algebra under the pointwise
multiplication . A review on how the algebra $({\cal A},.)$
captures all the differential geometry of $R^4$ can be found in
\cite{coq,madore,landi,varilly} .

It is known that we can make the algebra ${\cal A}$
non-commutative if we replace the dot product
 by the star prduct \cite{kon} . The pair $({\cal A}, *)$ is then describing a deformation $QR^4$ of space-time which
  will be
 taken
 by definition to be the quantum space-time . The $*$ product is defined for any two functions $f(x)$ and $g(x)$
 of ${\cal A}$ by \cite{witten}

\begin{equation}
f*g(x)=e^{\frac{i}{2}{C}^{{\mu}{\nu}}(x)\frac{\partial}{\partial{\xi}^{\mu}}\frac{\partial}{\partial{\eta}^{\nu}}}f(x+{\xi})g(x+{\eta})|_{{\xi}={\eta}=0}
\end{equation}
where ${C}^{{\mu}{\nu}}$ form a rank two tensor ${C}$ which in
general contains a symmetric as well as an antisymmetric
part\cite{peter} . It is assumed to be a function of $x$ of the
form
\begin{equation}
C^{{\mu}{\nu}}(x)={\chi}(x)({\theta}^{{\mu}{\nu}} +
 i a {\eta}^{{\mu}{\nu}})
\end{equation}
where ${\chi}(x)$ is some function of $x$ . ${\theta}$ is the
antisymmetric part and it is an $x$-independent tensor . $a$ is
as we will see the non-associativity parameter and it is
determined in terms of the tensor $\theta$ as follows . The
requirement that the star product $(2.1)$ is associative can be
expressed as the condition that $I=0$ where $I$ is given by :

\begin{equation}
I=(e^{ipx}*e^{ikx})*e^{ihx}-e^{ipx}*(e^{ikx}*e^{ihx}).
\end{equation}
$e^{ipx}$ are the generators of the algebra ${\cal A}$ written in
their bounded forms . Using the definition $(2.1)$ we can check
that
\begin{equation}
e^{ipx}*e^{ikx} =e^{-\frac{i}{2}pCk}e^{i(p+k)x}
\end{equation}
and therefore $(2.3)$ takes the form

\begin{eqnarray}
Ie^{-i(p+k+h)x}&=&e^{-\frac{1}{2}C^{{\mu}{\nu}}(x)h_{\nu}\frac{\partial}{\partial{{\xi}^{\mu}}}}[e^{-\frac{i}{2}pC(x+{\xi})k
+ i(p+k){\xi}}]|_{{\xi}=0}\nonumber\\
&-&e^{-\frac{1}{2}C^{{\nu}{\mu}}(x)p_{\nu}\frac{\partial}{\partial{{\xi}^{\mu}}}}[e^{-\frac{i}{2}kC(x+{\xi})h
+ i(k+h){\xi}}]|_{{\xi}=0}.
\end{eqnarray}

To see clearly what are the kind of conditions we need to ensure
that the equation $I=0$ is an identity , we first expand both
sides of $(2.5)$ in powers of $C$ and keep terms only up to the second
order . It will then read
\begin{equation}
I=\frac{i}{4}\bigg[C^{{\mu}{\nu}}h_{\nu}p{\frac{{\partial}C}{{\partial}x^{\mu}}}k
-C^{{\nu}{\mu}}p_{\nu}k{\frac{{\partial}C}{{\partial}x^{\mu}}}h\bigg].
\end{equation}
As we can clearly see the associativity of the star product at
this order is maintaned if and only
$C^{{\mu}{\nu}}{\partial}_{\mu} C=0 $ and $
C^{{\nu}{\mu}}{\partial}_{\mu} C=0 $ . The two consequences of
these two conditions are given by the equations $
a{\eta}^{{\mu}{\nu}}{\partial}_{\mu}{\chi}=0 $ and
${\theta}^{{\mu}{\nu}}{\partial}_{\mu}{\chi}=0$ . The first
equation is simply $a=0$ because the solution ${\chi}=constant$
will be discarded in this paper . The second equation in the
other hand means as we can simply check that the noncommutativity
matrix $\theta$ is singular , i.e $det{\theta}=0$ . We can aslo
check that the two above conditions are necessary and sufficient
to make the star product $(2.1)$ associative at all orders
because of the identities $
{\theta}^{{\mu}_{1}{\nu}_{1}}{\theta}^{{\mu}_{2}{\nu}_{2}}..{\theta}^{{\mu}_{n}{\nu}_{n}}{\partial}^{n}_{{\mu}_1,{\mu}_2,..,{\mu}_n}
{\chi}=0 $.

If we would like to avoid the singularity of the noncommutativity
matrix $\theta$ we have then to relax the requirement of
associativity . We can start by reducing the associativity of the
star product $(2.1)$ by imposing only one of the above two
conditions , say
\begin{eqnarray}
C^{{\mu}{\nu}}\frac{\partial C}{\partial{x^{\mu}}}&=&0\nonumber\\
&{\Longrightarrow}&\nonumber\\
C^{{\mu}{\nu}}\frac{\partial{\chi}}{\partial{x^{\mu}}}&=&0.
\end{eqnarray}
Before we analyze further this equation , we remark that this
condition on the tensor $C$ will lead to the identities
\begin{equation}
C^{{\mu}_{1}{\nu}_{1}}C^{{\mu}_{2}{\nu}_{2}}..C^{{\mu}_{n}{\nu}_{n}}{\partial}^{n}_{{\mu}_1,{\mu}_2,..,{\mu}_n}
{C^{{\alpha}{\beta}}}=0 .
\end{equation}
$(2.7)$ will also lead to the equation
\begin{equation}
C^{{\nu}{\mu}}\frac{\partial
C}{\partial{x^{\mu}}}=i{\partial}^{\nu}{\chi}^2\bigg[a{\theta} +
i a^2{\eta}\bigg].
\end{equation}
In order to have a very small amount of nonassociativity in the
theory we will assume that $a$ is a very small parameter in such a
way that only linear terms in $a$ are relevant . Putting $(2.7)$
and $(2.9)$ in $(2.5)$ will then give

\begin{equation}
I=\frac{ia}{2}(k{\theta}h)O(p,k+h,{\chi},{\partial}{\chi})e^{-\frac{i{\chi}}{2}k{\theta}h}e^{i(p+k+h)x},
\end{equation}
where $O$ is a function (which we will not write down explicitly
)of the momenta p,k,h and of ${\chi}$ and all of its derivatives
$\{{\partial}{\chi}\}$ . This function $O$ is such that it
vanishes identically if ${\partial}_{\mu}{\chi}=0$ . In other
words a trivial solution to the equation $I=0$ is
${\chi}=constant$ which we will discard in this paper . We would
like to determine ${\chi}$ from the requirement that the quantum
field theory which we will eventually write
 down on $QR^4$ is finite . So we will leave ${\chi}$ arbitrary at this stage .
 Clearly ${\chi}$ will be model depenedent and it can generally be put in the form

\begin{equation}
{\chi}(x)=\sum_{n=1}{\hbar}^n{\chi}_n(x)
\end{equation}
where we don't have a tree level term because by assumption this function will be entirely determined by the different
 infinities of the theory which are generally of higher orders in $\hbar$ . In other words
 the zero order is absent in $(2.11)$ because QFT's are usually finite at this order .

It is instructive to solve equation $(2.7)$ for ${\theta}$ in
terms of ${\chi}$ . We assume that
${\partial}_{\mu}{\chi}{\neq}0$ and rewrite the equation $(2.7)$
in the form $
C^{{\mu}{\nu}}{\partial}_{\mu}{\chi}={\lambda}e^{\nu} $  where
${\lambda}$ is a small number and $e$ is a four-vector given by
$(1,0,0,0)$ . Solving $(2.7)$ for ${\theta}$ will give
  the following equation

\begin{eqnarray}
\frac{idetC}{a^3 -\frac{a}{2}\sum_{{\mu},{\nu}{\neq}0}{\theta}_{{\mu}{\nu}}{\theta}^{{\mu}{\nu}}}&=&{\lambda}\frac{{\chi}^3}{{\partial}_{0}{\chi}}\nonumber\\
\frac{detC}{-a^2{\theta}^{0i} -
ia{\theta}_{i{\mu}}{\theta}^{0{\mu}}
+{\theta}^{jk}\sqrt{det{\theta}}}&=&{\lambda}\frac{{\chi}^3}{{\partial}_{i}{\chi}},
\end{eqnarray}
with
\begin{equation}
detC={\chi}^4[det{\theta} + a^4 -
\frac{a^2}{2}{\theta}_{{\mu}{\nu}}{\theta}^{{\mu}{\nu}}].
\end{equation}
$(ijk)$ are the even permutations of $(123)$ and $det{\theta}$ is
given by : $
det{\theta}=[\frac{1}{8}{\epsilon}_{{\mu}{\nu}{\alpha}{\beta}}{\theta}^{{\mu}{\nu}}{\theta}^{{\alpha}{\beta}}]^2
$ . The $4$ equations $(2.12)$ provide $4$ constraints on the
tensor $\theta$ which reduce at limit
${\lambda}{\longrightarrow}0$ to one constarint given by
\begin{equation}
det{\theta} =- a^4 +
\frac{a^2}{2}{\theta}_{{\mu}{\nu}}{\theta}^{{\mu}{\nu}}
\end{equation}
This is a generalization of the quantization conditions chosen in
\cite{dop} . This equation however can be thought of as giving
the nonassociativity parameter $a$ in terms of the
noncommutativity matrix ${\theta}$ . The solution is
\begin{equation}
a=\bigg[\frac{1}{4}{\theta}_{{\mu}{\nu}}{\theta}^{{\mu}{\nu}} -
\sqrt{(\frac{1}{4}{\theta}_{{\mu}{\nu}}{\theta}^{{\mu}{\nu}})^2 -
det{\theta}}\bigg]^{\frac{1}{2}}.
\end{equation}
 As we can see from the above analysis it is
necessary and sufficient to choose ${\theta}$ in such a way that
 $(2.15)$ is a very small number in order for the associativity
of the star product $(2.1)$ to be broken with the very small
amount given by $(2.10)$ .

Using the $*$ product $(2.1)$ we can define the Moyal bracket of
any two functions $f(x)$ and $g(x)$ by $\{f(x),g(x)\}=f*g(x) -
g*f(x)$ and in particular the Moyal bracket of two coordinate
functions is given by

\begin{equation}
\{x^{\mu},x^{\nu}\}=i{\chi}(x){\theta}^{{\mu}{\nu}}.
\end{equation}
For self-consistency this bracket should satisfy the Jacobi identity
\begin{equation}
\{x^{\beta},\{x^{\mu},x^{\nu}\}\}+\{x^{\nu},\{x^{\beta},x^{\mu}\}\}+\{x^{\mu},\{x^{\nu},x^{\beta}\}\}=0 ,
\end{equation}
but
\begin{equation}
\{x^{\beta},\{x^{\mu},x^{\nu}\}\}=-ia{\chi}({\partial}^{\beta}{\chi}){\theta}^{{\mu}{\nu}}.
\end{equation}
Clearly at the limit of associativity ($a{\longrightarrow}0$)
 , equation $(2.18)$ is simply zero and therefore $(2.17)$ holds . We
would like however to maintain Jacobi identity even for
$a{\neq}0$ . we then need to impose the following constraint on
${\theta}$
\begin{equation}
{\theta}^{{\alpha}{\beta}}{\theta}^{{\mu}{\nu}}+{\theta}^{{\alpha}{\nu}}{\theta}^{{\beta}{\mu}}+{\theta}^{{\alpha}{\mu}}{\theta}^{{\nu}{\beta}}=0.
\end{equation}
which will make $(2.17)$ an identity . A class of  solutions to
the equation $(2.19)$ can be given by those antisymmetric tensors
${\theta}$ such that
\begin{equation}
{\theta}^{{\mu}{\nu}}={a}^{\mu}_{\alpha}{a}^{\nu}_{\beta}{\theta}_0^{{\alpha}{\beta}}
\end{equation}
where $ {a}^{\mu}_{\alpha} $ are arbitrary real numbers , and ${\theta}_0$ is an antisymmetric tensor which satisfies

\begin{equation}
{\theta}_0^{{\mu}{\nu}}{\theta}_0^{{\alpha}{\beta}}=({\eta}^{{\mu}{\alpha}}{\eta}^{{\nu}{\beta}} - {\eta}^{{\mu}{\beta}}{\eta}^{{\nu}{\alpha}}).
\end{equation}
$(2.19)$ is the only constarint we need to impose on the tensors
$\theta$ in order to have both the associativity requirement in
the sense of $(2.10)$ and Jacobi identiy $(2.17)$ to be satisfied
. By requiring that $(2.16)$ should lead to a certain kind of
space-time uncertainty relations we can further restrict the
allowed antisymmetric tensors ${\theta}$ as we will see in the
next section .

%\begin{eqnarray}
%f*g(x)&=&f(x)g(x) + \frac{i}{2}{\chi}(x)({\theta}^{{\mu}{\nu}} + S^{{\mu}{\nu}%})\frac{\partial{f}}{\partial{x^{\mu}}}\frac{\partial{g}}{\partial{x^{\nu}}}\n%onumber\\
%& -& \frac{1}{8}{\chi}^2(x) ({\theta}^{{\mu}{\nu}} + S^{{\mu}{\nu}}) ({\theta}%^{{\alpha}{\beta}} + S^{{\alpha}{\beta}}) \frac{{\partial}^2{f}}{\partial{x^{\%mu}}\partial{x^{\alpha}}}\frac{{\partial}^2{g}}{\partial{x^{\nu}}\partial{x^{\%beta}}}.
%\end{eqnarray}
%The Moyal bracket of any two functions $f(x)$ and $g(x)$ in the other hand wil%l have the form

%\begin{equation}
%\{f,g\}= i{\chi}(x){\theta}^{{\mu}{\nu}} \frac{\partial{f}}{\partial{x^{\mu}}}%\frac{\partial{g}}{\partial{x^{\nu}}} - \frac{1}{2}{\chi}^2(x){\theta}^{{\alph%a}{\beta}}S^{{\mu}{\nu}} \frac{{\partial}^2{f}}{\partial{x^{\mu}}\partial{x^{\%alpha}}}\frac{{\partial}^2{g}}{\partial{x^{\nu}}\partial{x^{\beta}}}.
%\end{equation}

\vskip 5mm
\noindent {\bf 2.2 The Algebra $({\cal A},*)$}
\vskip 5mm
\noindent

A general element $f(x)$ of ${\cal A}$ will be
defined by

\begin{equation}
f(x)=\int \frac{d^4p}{(2{\pi})^4}\tilde{f}(p,\chi) e^{ipx}
\end{equation}
where $\tilde{f}$ is a smooth continuous function of the 4-vector
$p$ and of the fuzzyness function $\chi$ which satisfies
$\tilde{f}^{*}(-p,\chi)=\tilde{f}(p,\chi)$ . It is of the general
form $\tilde{f}(p,\chi)={\tilde{f}}_0(p,{\chi}) + a
{\tilde{f}}_1(p,{\chi}) $ . The $*$ product $(2.1)$ can then be
rewritten as

\begin{eqnarray}
f*g(x)&=&\int\frac{d^4p}{(2{\pi})^4}\frac{d^4k}{(2{\pi})^4}\bigg[\tilde{f}(p,\chi)\tilde{g}(k,\chi)e^{-\frac{i}{2}pCk}\nonumber\\
&+&a{\tilde{f}}(p,\chi){\frac{{\partial}{\tilde{g}(k,\chi)}}{{\partial}{\chi}}}O(p,k,\chi,{\partial}{\chi})\bigg]e^{i(p+k)x} .\nonumber\\
&=&\int\frac{d^4p}{(2{\pi})^4}\tilde{f}*\tilde{g}(p,\chi)e^{ipx}.
\end{eqnarray}
$O(p,k,\chi,{\partial}{\chi})$ is the function defined by the
equation $(2.10)$ . The Fourier transform
$\tilde{f}*\tilde{g}(p,\chi)={\tilde{f}*\tilde{g}}(p,\chi)_0+a{\tilde{f}*\tilde{g}}(p,\chi)_1$
is given in the other hand by

\begin{eqnarray}
{\tilde{f}*\tilde{g}}(p,\chi)&=&\int
\frac{d^4k}{(2{\pi})^4}\bigg[{\tilde{f}}(p-k,\chi){\tilde{g}}(k,\chi)e^{-\frac{i}{2}(p-k)Ck}\nonumber\\
&+&a{\tilde{f}}(p-k,\chi){\frac{{\partial}{{\tilde{g}}(k,\chi)}}{{\partial}{\chi}}}O(p-k,k,\chi,{\partial}{\chi})\bigg].
\end{eqnarray}
The function $\tilde{f}(p,\chi)$ can always be expanded as :
$\tilde{f}(p,\chi)=\sum_{n=0} a_n \bar{f}_{n}(\chi)\tilde{f}(p)$
which suggests that $(2.22)$ can be rewritten in the form
\cite{dop}

\begin{equation}
f(x)=\sum_{n=0}a_{n}f_{n}(x)
\end{equation}
where
\begin{equation}
f_{n}(x) = \bar{f}_{n}(\chi)\int
\frac{d^4p}{(2{\pi})^4}\tilde{f}_{n}(p) e^{ipx}.
\end{equation}
$f_{n}(x)$ are the generators of the algebra $({\cal A},*)$
written in a way which will allow us to see the classical limit
defined by ${\chi}{\longrightarrow}0$ . In this limit they must
generate the algebra $ ({\cal A},.)$ . Therefore the functions
$\bar{f}_{n}(\chi)$ are such that they tend to a constant when
${\chi}{\longrightarrow}0$ . This constant can always be chosen to
be $1$  .

\vskip 5mm
\noindent {\bf 2.3 Change of Generators Basis }
\vskip 5mm
\noindent

Finally we would like to rewrite $(2.16)$ in way which will be
more suitable for quantization . This will involve finding a basis
$z^{\mu}(x)$ for which  the Moyal bracket $\{z^{\mu},z^{\nu}\}$
is in the center of the algebra $({\cal A},*)$ , in other words
$\{x^{\alpha},\{z^{\mu},z^{\nu}\}\}=0$ . This is not the case for
the basis $x^{\mu}$ as we can see from equation $(2.18)$ . We then
must have $\{z^{\mu},z^{\nu}\}=i{\theta}^{{\mu}{\nu}}{\cal C}(x)$
where ${\cal C}(x)$ is any function of $x$ which does commute (in
the sense of Moyal bracket) with the elements of the algebra
$({\cal A},*)$ . To find such a basis we need first to find the
central elements ${\cal C}(x)$ of the algebra $({\cal A},*)$ . To
this end we first remark that by using the equation $(2.1)$ the
Moyal bracket of the generator $x^{\mu}$ with any function $f(x)$
is given by
\begin{equation}
\{x^{\mu},f\}=i{\chi}{\theta}^{{\mu}{\nu}}\frac{\partial{f}}{\partial{x^{\nu}}}.
\end{equation}
It is then clear that the only obvious solutions to the equation
$\{x^{\mu},f\}=0$ are the trivial ones , namely the constant
functions . However choosing the central element ${\cal C}(x)$ to
be a constant is not good because it will lead to a singular
basis at ${\chi}(x)=0$ which can be seen from the fact that the
Moyal bracket $\{z^{\mu},z^{\nu}\}$ at ${\chi}(x)=0$ will then not
vanish on the contrary to what happens to the Moyal bracket
$(2.16)$ which clearly vanishes at ${\chi}=0$ . So we must find at
least one central element which is not a constant function . The
only clear way to find such an element is to use perturbation
theory . We assume then that the quantum field theory which we
will write on $QR^4$ is relevant only up to the ${\hbar}^N$ order
. The function ${\chi}(x)$ will then take the form
\begin{equation}
{\chi}(x)=\sum_{n=1}^{N}{\hbar}^n{\chi}_n(x)
\end{equation}
and we would have that
\begin{equation}
{\chi}^{N+1}(x)=0.
\end{equation}
This last equation can be rewritten by using  equation $(2.27)$ as
\begin{equation}
\{x^{\mu},{\chi}^N\}=0,
\end{equation}
in other words ${\chi}^N$ is a central element of the algebra
${\cal A}$ in the ${\hbar}^N$ approximation . Actually any
combination of the order of ${\hbar}^N$ is central as it can be
seen from equations $(2.27)$ and $(2.29)$ . By choosing ${\cal
C}(\chi)={\chi}^N(x)$ , the Moyal bracket of any two coordinates
$z^{\mu}(x)$ and $z^{\nu}(x)$ will then read

\begin{equation}
\{z^{\mu},z^{\nu}\}=i{\chi}^N{\theta}^{{\mu}{\nu}}.
\end{equation}
$x^{\mu}$ and $z^{\mu}(x)$ give equivalent descriptions of the
algebra $({\cal A},*)$ and therefore the quantization of $(2.16)$
is equivalent to the quantization of $(2.31)$ . It is obvious
however that the quantization of $(2.31)$ is more straight forward
than the quantization of $(2.16)$ . The new basis $z^{\mu}(x)$ can
be found in terms of $x^{\mu}$ as follows . First we note that
for the purpose of finding $z^{\mu}$  it is sufficent to work up
to the second order in $C$ . the star product $(2.1)$ of any two
functions $f(x)$ and $g(x)$ will read up to this order
\begin{eqnarray}
f*g(x)&=&f(x)g(x) + \frac{i}{2}{\chi}(x)({\theta}^{{\mu}{\nu}} + i a {\eta}^{{\mu}{\nu}})\frac{\partial{f}}{\partial{x^{\mu}}}\frac{\partial{g}}{\partial{x^{\nu}}}\nonumber\\
& -& \frac{1}{8}{\chi}^2(x) ({\theta}^{{\mu}{\nu}} + i
a{\eta}^{{\mu}{\nu}}) ({\theta}^{{\alpha}{\beta}} + ia
{\eta}^{{\alpha}{\beta}})
\frac{{\partial}^2{f}}{\partial{x^{\mu}}\partial{x^{\alpha}}}\frac{{\partial}^2{g}}{\partial{x^{\nu}}\partial{x^{\beta}}},
\end{eqnarray}
and therefore the Moyal bracket of these two functions is

\begin{equation}
\{f,g\}= i{\chi}(x){\theta}^{{\mu}{\nu}}
\frac{\partial{f}}{\partial{x^{\mu}}}\frac{\partial{g}}{\partial{x^{\nu}}}
-
\frac{ia}{2}{\chi}^2(x){\theta}^{{\alpha}{\beta}}{\eta}^{{\mu}{\nu}}
\frac{{\partial}^2{f}}{\partial{x^{\mu}}\partial{x^{\alpha}}}\frac{{\partial}^2{g}}{\partial{x^{\nu}}\partial{x^{\beta}}}.
\end{equation}
In particular the Moyal bracket of the two coordinates $z^{\mu}(x)$ and $z^{\nu}(x)$ is given by

\begin{equation}
\{z^{\delta},z^{\sigma}\}= i{\chi}(x){\theta}^{{\mu}{\nu}}
\frac{\partial{z^{\delta}}}{\partial{x^{\mu}}}\frac{\partial{z^{\sigma}}}{\partial{x^{\nu}}}
-
\frac{ia}{2}{\chi}^2(x){\theta}^{{\alpha}{\beta}}{\eta}^{{\mu}{\nu}}
\frac{{\partial}^2{z^{\delta}}}{\partial{x^{\mu}}
\partial{x^{\alpha}}}\frac{{\partial}^2{z^{\sigma}}}{\partial{x^{\nu}}\partial{x^{\beta}}}.
\end{equation}
Comapring $(2.31)$ and $(2.34)$ will then give that
\begin{eqnarray}
{\theta}^{{\mu}{\nu}} \frac{\partial{z^{\delta}}}{\partial{x^{\mu}}}\frac{\partial{z^{\sigma}}}{\partial{x^{\nu}}}&=&{\chi}^{N-1}{\theta}^{{\delta}{\sigma}}\nonumber\\
{\Longrightarrow}
\frac{\partial{z^{\mu}}}{\partial{x^{\nu}}}&=&{\chi(x)}^{\frac{N-1}{2}}\eta^{\mu}_{\nu}.
\end{eqnarray}
$(2.35)$ define scaling transformations which depend on space-time
points . A more thorough study of these transformations will be
reported elsewhere . As we can clearly see the definition $(2.35)$
of the new basis $z^{\mu}$ in terms of $x^{\mu}$ will make the
quadratic term in $(2.34)$ vanishes , and for that matter all
terms which are higher orders in $C$ will also vanish. We would
like now to rewritte $(2.35)$ in a form which is better suited for
quantization . To this end we make use of the equation $(2.27)$
for the case where $f=z^{\nu}$ . We then obtain

\begin{equation}
\{x^{\mu},z^{\nu}\}=i{{\chi}(x)}^{\frac{N+1}{2}}{\theta}^{{\mu}{\nu}},
\end{equation}
where we have used $(2.35)$ . Equation $(2.36)$ is actually
$(2.35)$ only written in terms of Moyal bracket which under
quantization will go to the commutator
 as we will see . For the coordinates $z^{\mu}$ the Jacobi identity $\{z^{\mu},\{z^{\nu},z^{\alpha}\}\} +
 \{z^{\alpha},\{z^{\mu},z^{\nu}\}\} + \{z^{\nu},\{z^{\alpha},z^{\mu}\}\} = 0$ trivially follows from $(2.31)$ .

By using the equation $(2.33)$ we can find that the Moyal bracket
of the generator $z^{\mu}$ with any function $f$ of ${\cal A}$
can be written as
\begin{equation}
\{z^{\mu},f\}=i{\chi}^N{\theta}^{{\mu}{\nu}}\frac{\partial{f}}{\partial{z^{\nu}}},
\end{equation}
where we have made use of $(2.35)$ . The Moyal brackets $(2.31)$
and $(2.37)$ do clearly correspond to the star product

\begin{equation}
f*g(z)=e^{\frac{i}{2}{D}^{{\mu}{\nu}}(z)\frac{\partial}{\partial{\xi}^{\mu}}\frac{\partial}{\partial{\eta}^{\nu}}}f(z+{\xi})g(z+{\eta})|_{{\xi}={\eta}=0}
\end{equation}
where now  $D^{{\mu}{\nu}}(z)={\chi}^N( {\theta}^{{\mu}{\nu}} +
ia{\eta}^{{\mu}{\nu}} ) $ . This star product however is
completely equivalent to $(2.1)$ . It is simply the star product
$(2.1)$ written in the basis $z^{\mu}$  . A general element of
the algebra $({\cal A},*)$ will be written in this basis as

\begin{equation}
f(z)=\int \frac{d^4p}{(2{\pi})^4}\tilde{f}(p) e^{ipz}
\end{equation}
where $\tilde{f}(p)={\tilde{f}}_0(p)+a{\tilde{f}}_1(p)$ . The
star product $(2.38)$ will then have the form
\begin{eqnarray}
f*g(z)&=&\int\frac{d^4p}{(2{\pi})^4}\frac{d^4k}{(2{\pi})^4}\tilde{f}(p)\tilde{g}(k)e^{-\frac{i}{2}pDk}e^{i(p+k)z} .\nonumber\\
&=&\int\frac{d^4p}{(2{\pi})^4}\tilde{f}*\tilde{g}(p)e^{ipz}.
\end{eqnarray}
where $\tilde{f}*\tilde{g}(p)$ is given by
\begin{equation}
\tilde{f}*\tilde{g}(p)=\int\frac{d^4k}{(2{\pi})^4}\tilde{f}(p-k)\tilde{g}(k)e^{-\frac{i}{2}(p-k)Dk}.
\end{equation}
In this case $\tilde{f}*\tilde{g}(p)$ is a function only of
${\chi}^N$ and not of ${\chi}$ . However ${\chi}^N$ is simply a
constant in the ${\hbar}^n$ approximation and therefore $(2.40)$
is of the same form as $(2.39)$ .

 \vskip 5mm \sxn{Quantum Space-Time}
\vskip 5mm \noindent {\bf 3.1 Quantization} \vskip 5mm \noindent
We will now show that the algebra $({\cal A},*)$ does really
describe a quantum space-time . In other words $QR^4$ is a
space-time we obtain by quantizating $R^4$ in the following way .
First of all we assume that the quantization of $R^4$ is
completely equivalent to the quantization of its underlying
algebra $({\cal A},.)$ \cite{madore,landi}. Then in analogy with
Quantum Mechanics we will quantize $({\cal A},.)$ by the usual
quantization prescription of replacing the coordinate functions
$x^{\mu}$ by the coordinate operators $X^{\mu}$ so that the
algebra of functions $({\cal A},.)$ is mapped to an algebra of
operators $(A,{\times})$ \cite{dop}. If this algebra of operators
$(A,\times)$ is to be describing the quantum space-time $QR^4$ it
must be constructed in such a way that it will be homomorphic to
$({\cal A},*)$ . In other words we must construct a homomorphism
${\cal X}$ from $(A,\times)$ to $({\cal A},*)$ which will map any
element $F(X)$ of $A$ to the element $(2.22)$ of $({\cal A},*)$ in
such a way that the operator product $F(X){\times}G(X)$ is mapped
to the star product $(2.23)$ . We would then have
\begin{equation}
F(X){\longrightarrow}{\cal X}(F(X))=f(x)
\end{equation}
together with
\begin{equation}
F(X){\times}G(X){\longrightarrow}{\cal X}(F(X){\times}G(X))=f*g(x)
\end{equation}
where $g(x)$ is the image of the operator $G(X)$ . In particular
from $(3.1)$ the coordinate operators $X^{\mu}$ are mapped to the
coordinate functions $x^{\mu}$ and from $(3.2)$ the Moyal
bracket$\{f,g\}$ is mapped to the commutator
$[F,G]_{\times}=F{\times}G-G{\times}F$\cite{merkulov}. As we will
see the homomorphism ${\cal X}$ has no non trivial kernals and
therefore the arrows in $(3.1)$ and $(3.2)$ can go the other way .

The product ${\times}$ which we will call the nonassociative
operator product cannot be the ordinary dot product of operators
because it is clear from the definition $(3.2)$ that $\times$ is
nonassociative whereas the dot product of operators is trivially
an associative product . We can assume however that it will reduce
at the limit of $a{\longrightarrow}0$ to the ordinary dot product
of operators . The difference ${\Delta}$ between the
nonassociative product ${\times}$ and the ordinary dot product is
of the order of $a$ and it is given by

\begin{equation}
{\Delta}(F,G)=\frac{F{\times}G - F.G}{a}
\end{equation}
where $F.G$ is defined by
\begin{equation}
{\cal X}(F(X).G(X))=Lim_{a{\longrightarrow}0}f*g(x)
\end{equation}

The first step in constructing this homomorphism ${\cal X}$ is to
impose on the coordinate operators $X^{\mu}$ commutation
relations which are of the same form as $(2.16)$  . We then have

\begin{equation}
[X^{\mu},X^{\nu}]_{\times}=iR{\theta}^{{\mu}{\nu}}
\end{equation}
where $R$ is an operator defined by
\begin{equation}
{\cal X}(R)=\chi(x).
\end{equation}
In terms of the ordinary commutator , equation $(3.5)$ will
simply read
\begin{equation}
[X^{\mu},X^{\nu}]=iR{\theta}^{{\mu}{\nu}}.
\end{equation}
The contribution
${\Delta}(X^{\mu},X^{\nu})-{\Delta}(X^{\nu},X^{\mu})$ to this
commutator is identically zero because
${\Delta}(X^{\mu},X^{\nu})=-\frac{R}{2}{\eta}^{{\mu}{\nu}}$ .

The operator $R$ clearly does not commute with $X^{\mu}$ because
\begin{equation}
[R,X^{\mu}]_{\times}=R^{\mu}
\end{equation}
where $R^{\mu}$ are the elements of the algebra $A$ mapped to $\{{\chi},x^{\mu}\}$ , i.e
\begin{equation}
{\cal X}(
R^{\mu})=\{{\chi},x^{\mu}\}=-i{\chi}{\theta}^{{\mu}{\nu}}{\partial}_{\nu}{\chi}
\end{equation}
The equation $(3.8)$ will simply mean that the Jacobi identity
\begin{equation}
[X^{\mu},[X^{\nu},X^{\alpha}]_{\times}]_{\times}+[X^{\alpha},[X^{\mu},X^{\nu}]_{\times}]_{\times}+[X^{\nu},[X^{\alpha},X^{\mu}]_{\times}]_{\times}=0
\end{equation}
is not satisfied unless we choose ${\theta}$ to satisfy $(2.19)$ .

In general the commutator of the generator $X^{\mu}$ with any element $F(X)$ of the algebra $A$ is found to be
\begin{equation}
[X^{\mu},F]_{\times}={\Delta}F
\end{equation}
where by using $(3.1)$ and $(3.2)$ , ${\Delta}F$ is the operator
in $A$ mapped to $\{x^{\mu},f\}$ , i.e
\begin{equation}
{\cal X}({\Delta}F)=\{x^{\mu},f\}.
\end{equation}
It is clear from this equation that the central elements of the
algebra $A$ are either those operators which are mapped to the
constant functions or the operator $O$ which is mapped to
${\chi}^N$ . The operators mapped to the constant functions are
clearly multiples of the identity operator ${\bf 1}$ . The
operator $O$ in the other hand is $R^N$ which can be seen as
follows . By using equation $(2.32)$ we can prove that in the
${\hbar}^N$ approximation we have that
${\chi}*({\chi}*({\chi}*({\chi}..*({\chi}*{\chi})))..)={\chi}^{N}$
where we have $N$ factors in the product . This equation will
become under quantization $R^N + a
\sum_{m=0}^{N-2}R^m{\Delta}(R,R^{N-m-1})=O$ . However by using
the definition $(3.3)$ of ${\Delta}$ , one can check that in the
${\hbar}^N$ approximation the second term in the expression of
$O$ is of the order of ${\hbar}^{N+1}$ and therefore $O=R^N$ .
The generators $X^{\mu}$ will then commute with $R^{N}$ , i.e
\begin{equation}
[X^{\mu},R^{N}]_{\times}=0.
\end{equation}
In general $X^{\mu}$ will commute with any element of $A$ which is of the order of ${\hbar}^N$ .

The fact that $R$ does not commute with the algebra $A$ makes the
definition $(3.5)$ of quantum space-time not very useful when we
try to construct explicitly the homomorphism ${\cal X}$ . To see
this more clearly we first note that general elements $F(X)$ of
the algebra $A$ are of the form
\begin{equation}
F(X)=\int \frac{d^4p}{(2{\pi})^4}[\tilde{F}(p,R) e^{ipX} +
e^{-ipX}\tilde{F}^{+}(p,R)]
\end{equation}
The nonassociative product of any two such elements $F(X)$ and
$G(X)$ will involve four different terms because $R$ dose not
commute with $e^{(ipX)}$ .
 So there is no an obvious way on how to map $F(X)$ given by $(3.14)$ to $f(x)$ given by
  $(2.22)$ or for that matter how to map $F(X){\times}G(X)$ to the star product $f*g$ .

For the purpose of quantization a better definition of quantum
space-time $QR^4$ is such that the commutators of the generators
are in the center of the algebra . We need then to find a basis
$Z^{\mu}$ for which we have the commutators
$[Z^{\mu},Z^{\nu}]_{\times}=i{\theta}^{{\mu}{\nu}}{\cal C}$ where
${\cal C}$ is a central element of the algebra $A$ . If $Z^{\mu}$
is the operator in $A$ mapped to the coordinate function $z^{\mu}$
introduced in $(2.31)$ then ${\cal C}$ will be simply given by
$R^N$ . We would then have
\begin{equation}
[Z^{\mu},Z^{\nu}]_{\times}=i{\theta}^{{\mu}{\nu}}R^N.
\end{equation}
The ordinary commutator will also be given by a similar equation
$[Z^{\mu},Z^{\nu}]=i{\theta}^{{\mu}{\nu}}R^N $ because of the
fact that
${\Delta}(Z^{\mu},Z^{\nu})=-\frac{R^N}{2}{\eta}^{{\mu}{\nu}}$ .

The definition of the operators $Z^{\mu}$ in terms of $X^{\mu}$
can be given by the equation
\begin{equation}
[X^{\mu},Z^{\nu}]_{\times}=i{\theta}^{{\mu}{\nu}}R_{\times}^{\frac{N+1}{2}},
\end{equation}
with
\begin{equation}
{\cal X}(R_{\times}^{\frac{N+1}{2}})={\chi}^{\frac{N+1}{2}},
\end{equation}
 where We clearly have used the requirement that this equation
should be mapped to $(2.36)$ .

The coordinate operators $Z^{\mu}$ are clearly unbounded and one
would like to work with bounded operators . We will therefore
consider instead the operators $e^{ipZ}$ as the generators of the
algebra $A$ . A general element $F(Z)$ of $A$ will be defined by

\begin{equation}
F(Z)=\int \frac{d^4p}{(2{\pi})^4}\tilde{F}(p) e^{ipZ}
\end{equation}
$\tilde{F}$ is a smooth continuous function of the $4$-vector $p$
which must satisfy $\tilde{F}^{+}(-p)=\tilde{F}(p)$ in order for
$F(Z)$ to be hermitian .

The product of any two elements $F(Z)$ and $G(Z)$ of $A$ can be
found to be

\begin{eqnarray}
F(Z){\times}G(Z)&=&\int\frac{d^4p}{(2{\pi})^4}\frac{d^4k}{(2{\pi})^4}\tilde{F}(p)\tilde{G}(k)e^{-\frac{iR^N}{2}p{\theta}k}e^{i(p+k)Z}\nonumber\\
&+&a\int\frac{d^4p}{(2{\pi})^4}\frac{d^4k}{(2{\pi})^4}\tilde{F}(p)\tilde{G}(k){\Delta}(e^{ipZ},e^{ikZ}).
\end{eqnarray}
where we have made use of Weyl formula :

\begin{equation}
e^{ipZ}e^{ikZ}=e^{-i\frac{R^N}{2}p{\theta}k}e^{i(p+k)Z}.
\end{equation}

 \vskip 5mm \noindent {\bf 3.2 Coherent States} \vskip 5mm
\noindent

Until now we did not define the homomorphism ${\cal X}$ explicitly
and once this is done the quantization of $R^4$ will be completed
. We claim that ${\cal X}(F)$ is defined as the map taking $F$ to
its diagonal matrix element in the coherent states basis $|x>$
\cite{peter,chms,cdp} . If we are working in the basis $(3.15)$ instead of
$(3.5)$ then ${\cal X}(F)$ is defined as the map taking $F$ to
its diagonal matrix element in the coherent states basis $|z>$ .
In order to define ${\cal X}$ we need first to introduce the
coherent states basis $|z>$ . We start by performing a
coordinates transformation to bring ${\theta}$ to the standard
form $B$ given by \cite{bars}

\begin{equation}
{B}=a\left( \begin{array}{cc}
                                      i{\sigma}_{2}  & 0\\
                                     0  & i{\sigma}_{2}
                                    \end{array}
                                     \right).
\end{equation}
Where ${\sigma}_2$ is the pauli matrix . ${\theta}$ and $B$ are
related by $B = {\Lambda} {\theta}{\Lambda}^{T}$ where
${\Lambda}$ is an $SO(4)$ transformation . Equation $(3.15)$
becomes under this transformation

\begin{equation}
[Q^{\mu},Q^{\nu}]=iR^NB^{{\mu}{\nu}}.
\end{equation}
Where $Q^{\mu}$ are the new coordinate operators and they are given in terms of $Z^{\mu}$ by the equations
 $ Q^{\mu}={{\Lambda}^{\mu}}_{\nu} Z^{\nu}$ . The only non vanishing commutation relations in $(3.22)$ are $[Q^{0},Q^{1}]=[Q^{2},Q^{3}]=iaR^N$
 and as we can see we have two commuting sets of conjugate variables $(Q^0,Q^1)$ and $(Q^2,Q^3)$ .
 Therefore we need to introduce only two commuting sets of creation and annhilation operators $(a,a^{+})$
and $(b,b^{+})$ . These creation and annhilation operators are defined by

\begin{eqnarray}
a&=&\frac{1}{\sqrt{2aR^N}}(Q^0+iQ^1)\nonumber\\
b&=&\frac{1}{\sqrt{2aR^N}}(Q^{2}+iQ^{3}).
\end{eqnarray}
The commutation relations $(3.22)$ in terms of these creation and
annhilation operators read $[a,a^{+}]=[b,b^{+}]=1$ . A state
$|n>$ ($n{\in}Z_{+}$) of the harmonic oscillator $(a,a^+)$ is
defined by $a^{+}|n>=\sqrt{n+1}|n+1>$ and $a|n>=\sqrt{n}|n-1>$ .
In the same way a state $|m>$ ($m{\in}Z_{+}$) of the harmonic
oscillator $(b,b^+)$ is defined by $b^{+}|m>=\sqrt{m+1}|m+1>$ and
$b|m>=\sqrt{m}|m-1>$ . Following \cite{klauder} we can then
introduce the coherent states $|q^0q^1>$ and $|q^2q^3>$ defined
by the equations

\begin{eqnarray}
|q^0q^1>&=&e^{-\frac{{q^0}^2+{q^1}^2}{4aR^N}}\sum_{n=0}^{\infty}\frac{(q^0 + i q^1)^n}{(2aR^N)^{\frac{n}{2}}\sqrt{n!}}|n>\nonumber\\
|q^2q^3>&=&e^{-\frac{{q^2}^2 +
{q^3}^2}{4aR^N}}\sum_{m=0}^{\infty}\frac{(q^2 + i
q^3)^m}{(2aR^N)^{\frac{m}{2}}\sqrt{m!}}|m>.
\end{eqnarray}
These coherent states can also be written as

\begin{eqnarray}
|q^0q^1>&=&U(q^0,q^1)|0>\nonumber\\
|q^2q^3>&=&U(q^2,q^3)|0>.
\end{eqnarray}
Where the operators $U(q^0,q^1)$ and $U(q^2,q^3)$ are given by

\begin{eqnarray}
U(q^0,q^1)&=&exp(\frac{i}{aR^N}(q^1Q^0 - q^0Q^1))\nonumber\\
U(q^2,q^3)&=&exp(\frac{i}{aR^N}(q^3Q^2 - q^2Q^3)).
\end{eqnarray}
These operators have the property that

\begin{eqnarray}
U^{-1}(q^0,q^1)({\alpha}Q^0 +{\beta}Q^1)U(q^0,q^1)&=&{\alpha}(Q^0 + q^0) +{\beta}(Q^1 + q^1)\nonumber\\
U^{-1}(q^2,q^3)({\alpha}Q^2 +{\beta}Q^3)U(q^2,q^3)&=&{\alpha}(Q^2 + q^2) +{\beta}(Q^3 + q^3).
\end{eqnarray}
where ${\alpha}$ and ${\beta}$ are arbitrary complex numbers .
This property simply means that the effect of $U(q^0,q^1)$ or
$U(q^2,q^3)$ on the operators $Q^0$ and $Q^1$ or $Q^2$ and $Q^3$
is to translate them by the c-numbers $q^0$ and $q^1$ or $q^2$
and $q^3$ respectively . The operators $U(q^0,q^1)$ and
$U(q^2,q^3)$ are therefore called translation operators . Finally
a general coherent state of the theory is clearly given by

\begin{equation}
|q>=|q^0q^1>|q^2q^3>=U(q^0,q^1)U(q^2,q^3)|0>|0>.
\end{equation}
Using the above structure we can then show the identity
\begin{equation}
<q|e^{ipQ}|q>=e^{ipq}e^{-\frac{aR^N}{4}p^2}.
\end{equation}
The proof goes as follows

\begin{eqnarray}
<q|e^{ipQ}|q>&=&<q^0q^1|e^{i(p_0Q^0 +p_1Q^1)}|q^0q^1><q^2q^3|e^{i(p_2Q^2 +p_3Q^3)}|q^2q^3>\nonumber\\
&=&<0|e^{A}e^{B}e^{-A}|0><0|e^{C}e^{D}e^{-C}|0>.
\end{eqnarray}
Where $A=-\frac{i}{aR^N}(q^1Q^0 - q^0Q^1)$ , $B=i(p_0Q^0 +
p_1Q^1)$ , $C=-\frac{i}{aR^N}(q^3Q^2 - q^2Q^3)$ and $D=i(p_2Q^2 +
p_3Q^3)$ . By using Weyl formula $(3.20)$ we can then compute that
$exp{(A)}exp{(B)}exp{(-A)}=exp{([A,B])}exp{(B)}$ . However
$[A,B]=i(p_0q^0 + p_1q^1)$ and therefore
$<0|exp{(A)}exp{(B)}exp{(-A)}|0>=exp{i(p_0q^0 +
p_1q^1)}<0|exp{(B)}|0> $ . Using Weyl formula again we get that
$exp{(B)}=exp{(-{\xi}^{*}a^{+} +
{\xi}a)}=exp{(-\frac{|{\xi}|^2}{2})}
exp{(-{\xi}^{*}a^{+})}exp{({\xi}a)}$ where ${\xi}$ is given by
${\xi}=\sqrt{\frac{aR^N}{2}}(ip_0+p_1)$ . The final result is
$<0|exp{(A)}exp{(B)}exp{(-A)}|0>=exp{(i(p_0q^0 +
p_1q^1))}exp{(-\frac{aR^N}{4}(p_0^2 + p_1^2))} $ . Similar
calculation will give that
$<0|exp{(C)}exp{(D)}exp{(-C)}>=exp{(i(p_2q^2 +
p_3q^3))}exp{(-\frac{aR^N}{4}(p_2^2 + p_3^2))}$ . All of this put
together gives $(3.29)$ .
 However the formula $(3.29)$ is clearly valid in any other basis and not only in the basis $(3.22)$ .
Rotating back to the basis $(3.15)$ will then give
\begin{equation}
<z|e^{ipZ}|z>=e^{ipz}e^{-\frac{a{\chi}^N}{4}p^2},
\end{equation}
where it is understood that ${\cal X}^N $ is the eigenvalue of the
operator $R^N$ on the coherent state  $|z>$ defined by
$|z>=U({\Lambda}^{-1})|q>$ . $(3.31)$ is the basic identity
needed in defining the map ${\cal X}$ . To show this we rewrite
$(3.31)$ in the following way
\begin{equation}
e^{-\frac{a{\chi}^N}{4}{\frac{\partial}{{\partial}z^{\mu}}}{\frac{\partial}{{\partial}z_{\mu}}}}(<z|e^{ipZ}|z>)=e^{ipz}.
\end{equation}
We note that at at the limit of $p{\longrightarrow}0$ this
identity takes the form

\begin{equation}
<z|Z^{\mu}|z>=z^{\mu}.
\end{equation}
Equation$(3.32)$ suggests that we define the homomorphism ${\cal
X}$ by

\begin{equation}
F(Z){\longrightarrow}{\cal
X}(F(Z))=e^{-\frac{a{\chi}^N}{4}{\frac{\partial}{{\partial}z^{\mu}}}{\frac{\partial}{{\partial}z_{\mu}}}}(<z|F(Z)|z>)=f(z).
\end{equation}
Now putting $(2.39)$ and $(3.18)$ in $(3.34)$ and using $(3.31)$
we get that $ \tilde{f}(p)=\tilde{F}(p) $ which simply means that
${\cal X}$ has no non trivial kernals \cite{klauder}. The
homomorphism ${\cal X}$ needs also to satify the requirement

\begin{equation}
F{\times}G(Z){\longrightarrow}{\cal
X}(F{\times}G(Z))=e^{-\frac{a{\chi}^N}{4}{\frac{\partial}{{\partial}z^{\mu}}}{\frac{\partial}{{\partial}z_{\mu}}}}(<z|F{\times}G(Z)|z>)=f*g(z),
\end{equation}
which can be checked by putting  $(2.40)$ and $(3.19)$ in this
last equation and using again $(3.31)$ .

\vskip 5mm \noindent {\bf 3.3 Uncertainty Relations }\vskip 5mm
\noindent

A class of solutions to the the condition $(2.14)$ which was
found to be the necessary and sufficient condition for the
associativity to hold approximately in the sense of $(2.10)$ can
be given by
\begin{eqnarray}
-\frac{1}{a^4}det{\theta} &\equiv &\frac{1}{a^4}(\vec{e}.\vec{b})^2=cosh^2{\alpha}\nonumber\\
-\frac{1}{2a^2}{\theta}_{{\mu}{\nu}}{\theta}^{{\mu}{\nu}}&\equiv&\frac{1}{a^2}({\vec{e}}^2
-{\vec{b}}^2)=sinh^2{\alpha} ,
\end{eqnarray}
where $\vec{e}$ and $\vec{b}$ are defined by
\begin{equation}
\theta=\left( \begin{array}{cccc}
                                     0  & -ie_1 &-ie_2&-ie_3\\
                                     ie_1  & 0 &b_3&-b_2\\
                                     ie_2  &-b_3 &0&b_1\\
                                     ie_3 &b_2&-b_1&0
\end{array}\right),
\end{equation}
and ${\alpha}$ is a real number which can be taken to be a
function of $a$  . The value ${\alpha}=0$ corresponds to the case
considered in \cite{dop} . From the above two equations $(3.36)$
we can find that
\begin{equation}
e^2{\geq}b^2{\geq}a^2
\end{equation}
We would like now that the commutation relations $(3.5)$ lead to a
certain space-time uncertainty relations . This will (in
principle) further restrict the allowed antisymmetric tensors
${\theta}$ .  Using the basic identity of quantum mechanics  :
${\Delta}a^2 {\Delta}b^2{\geq}\frac{1}{4}|<[A,B]>|^2 $ where
${\Delta}a^2=<{\Delta}A^2>=<A^2> - <A>^2$ the space-time
uncertainty relations are

\begin{eqnarray}
({\Delta}x^{\mu})^2({\Delta}x^{\nu})^2&{\geq}&\frac{1}{4}|<[X^{\mu},X^{\nu}]>|^2=\frac{<R>^2}{4}|{\theta}^{{\mu}{\nu}}|^2\nonumber\\
&{\Longrightarrow}&\nonumber\\
({\Delta}x^{0})^2\sum_{{i=1}^{i=3}}({\Delta}x^{i})^2&{\geq}&\frac{<R>^2}{4}\vec{e}^2\nonumber\\
&and&\nonumber\\
\sum_{1{\leq}i<j{\leq}3}({\Delta}x^{i})^2({\Delta}x^{j})^2&{\geq}&\frac{<R>^2}{4}\vec{b}^2.
\end{eqnarray}
By using the facts $(\sum{\Delta}x^i)^2{\geq}\sum({\Delta}x^i)^2$ ,
 $(\sum_{i<j}{\Delta}x^i{\Delta}x^j)^2{\geq}\sum_{i<j}({\Delta}x^i{\Delta}x^j)^2$  and the equation $(3.38)$
 the above uncertainty relations will take the form

\begin{eqnarray}
{\Delta}x^0.\sum_{i=1}^{i=3}{\Delta}x^{i}&{\geq}&\frac{{\lambda}}{2}\nonumber\\
\sum_{1{\leq}i<j{\leq}3}{\Delta}x^i{\Delta}x^j&{\geq}&\frac{{\lambda}}{2}
.
\end{eqnarray}
where ${\lambda}=a<R>$ . These are the same uncertainty relations which were derived in \cite{dop} . We can conclude from
the relations $(3.40)$ that quantum space-time has a cellular
structure . The minimal volume ( the volume of one cell ) is $
(\sqrt{2{\pi}{\lambda}})^4 $ and therefore a finite volume $V$ of
quantum space-time contains $V/(\sqrt{2{\pi}{\lambda}})^4 $ states . An
estimation of the fuzziness of space-time would determine or at
least give a bound on ${\lambda}$ which will restrict further the
allowed tensors ${\theta}$ .

\vskip 5mm \sxn{Quantum Field Theories on $QR^4$}

\noindent {\bf 4.1 The Dirac Operator} \vskip 5mm \noindent Before
we try to write action integrals on a given space we need always
to define first the Dirac operator on it . This Dirac operator
will provide the notion of derivations on this space and by
constructing it we would have basically constructed Connes triplet
associated to this space \cite{connes} . For $QR^4$
 this triplet consists of a representation ${\Pi}(A)$ of the algebra $A$ underlying the quantum space-time in some Hilbert
 space  , the  Dirac operator $D$ and the Hilbert space $H$ on which it acts
 . In the last section we have already constructed the representation ${\Pi}(A)$ in terms of the coherent states basis $|x>$ .
 The corresponding Dirac operator in the other hand  will be defined
 by \cite{sachin}
\begin{equation}
\int d^4x
e^{-\frac{a{\chi}}{4}{\frac{\partial}{{\partial}x^{\mu}}}{\frac{\partial}{{\partial}x_{\mu}}}}(<x|[D,\Phi]_{\times}{\times}[D,\Phi]_{\times}|x>)=\int
d^4x {\partial}_{\mu}\phi*{\partial}^{\mu}\phi,
\end{equation}
where $\phi$ is any element of the algebra ${\cal A}$ and $\Phi$ is its corresponding operator in $A$ .
 Clearly the ordinary Dirac operator ${\cal D}$ on $R^4$ given by ${\cal D}={\gamma}^{\mu}{\partial}_{\mu}$ where
 $\{{\gamma}^{\mu}\}$ is the Clifford algebra satsifying $\{{\gamma}^{\mu},{\gamma}^{\nu}\}=2{\eta}^{{\mu}{\nu}}$ ,
  will satisfy $(4.1)$ in the limit ${\theta}{\longrightarrow}0$ . In other words it will satisfy the equation $Tr[{\cal D},\Phi][{\cal D},\Phi]=\int
d^4x {\partial}_{\mu}\phi{\partial}^{\mu}\phi $ .
  It is reasonable to assume that this Clifford algebra will not get modified under
   quantization of space-time so that we can write $D$ as
\begin{equation}
   D={\gamma}^{\mu}D_{{\mu}}+ aF .
\end{equation}
   This assumption can be justified by the fact that the ${\gamma}'s$ are
   not elements of the algebra ${\cal A}$ and therefore quantizing the algebra
    will not quantize them . $F$ in $(4.2)$ is a connection arising from the nonassociativity of the underlying
    algebra $({\cal A},*)$ and it is defined such that $(4.1)$
    takes the form
\begin{equation}
\int d^4x
e^{-\frac{a{\chi}}{4}{\frac{\partial}{{\partial}x^{\mu}}}{\frac{\partial}{{\partial}x_{\mu}}}}([D_{\mu},\Phi]_{\times}{\times}[D^{\mu},\Phi]_{\times})=\int
d^4x{\partial}_{\mu}\phi*{\partial}^{\mu}\phi.
\end{equation}
Comparing $(4.1)$ and $(4.3)$ we can find that $F$ should satisfy
the condition
\begin{eqnarray}
Tr\bigg[{\gamma}^{\mu}[D_{\mu},{\Phi}][F,\Phi] +
\frac{i}{8}{\sigma}^{{\mu}{\nu}}{\Delta}([D_{\mu},\Phi],[D_{\nu},\Phi])\bigg]&=&\nonumber\\-\frac{1}{8}({\eta}^{{\mu}{\nu}}-{\gamma}^{\mu}{\gamma}^{\nu})\int
d^4x{\chi}
{\partial}_{\alpha}{\partial}^{\alpha}(<x|[D_{\mu},\Phi][D_{\nu},\Phi]|x>),
\end{eqnarray}
where
$\frac{i}{2}{\sigma}^{{\mu}{\nu}}=[{\gamma}^{\mu},{\gamma}^{\nu}]$
. A trivial solution to the equation $(4.4)$ is given by
\begin{eqnarray}
[F,{\Phi}]&=&-\frac{i}{8}[{\gamma}^{\alpha}D_{\alpha},\Phi]^{-1}{\sigma}^{{\mu}{\nu}}{\Delta}([D_{\mu},\Phi],[D_{\nu},\Phi])
- [{\gamma}^{\alpha}D_{\alpha},\Phi]^{-1}F_0\nonumber\\
where\nonumber\\
<x|F_0|x>&=&\frac{1}{8}({\eta}^{{\mu}{\nu}}-{\gamma}^{\mu}{\gamma}^{\nu}){\chi}{\partial}_{\alpha}{\partial}^{\alpha}(<x|[D_{\mu},\Phi][D_{\nu},\Phi]|x>).
\end{eqnarray}

$D_{\mu}$ are by definition the quantum derivations on $QR^4$ and
they are given by
\begin{equation}
e^{-\frac{a{\chi}}{4}{\frac{\partial}{{\partial}x^{\mu}}}{\frac{\partial}{{\partial}x_{\mu}}}}(<x|[D_{\mu},\Phi]_{\times}|x>)={\partial}_{\mu}{\phi}.
\end{equation}
By $(3.34)$ and $(3.35)$ , equation $(4.6)$ satisfies $(4.3)$
trivially . To find the quantum derivations $D_{\mu}$ we first
have to reexpress the classical derivations in terms of the Moyal
bracket and the star product introduced in section $1$ and once
this is done the transition to the quantum derivations is quite
straightforward . It simply consists of replacing the Moyal
bracket by the commutator $[,]_{\times}$ and the star product by
the nonassociative operator product ${\times}$ as explained in the
last section . By using Moyal bracket $(2.33)$ any arbitrary
vector field ${\cal L}^{\mu}$ should satisfy

\begin{equation}
\{{\cal
L}^{\mu},\phi\}=i{\chi}{\theta}^{{\alpha}{\beta}}{\partial}_{\alpha}{\cal
L}^{\mu}{\partial}_{\beta}\phi,
\end{equation}
where $\phi$ is any element of the algebra ${\cal A}$ . It is
clear that we have to assume that ${\cal L}^{\mu}$ is of the
order of ${\hbar}^{N-1}$ in order to have only the term written
in equation $(4.7)$ . This vector ${\cal L}^{\mu}$ in the other
hand will be defined by

\begin{equation}
\{{\cal
L}^{\mu},\phi\}=i{\chi}^N{\theta}^{{\mu}{\beta}}{\partial}_{\beta}\phi.
\end{equation}
Comparing $(4.7)$ and $(4.8)$ we get that
\begin{eqnarray}
{\partial}_{\alpha}{\cal L}^{\mu}&=&{\chi}^{N-1}{\eta}_{\alpha}^{\mu}\nonumber\\
&{\Longrightarrow}&\nonumber\\
 {\cal L}^{\mu}&=& \int {\chi}^{N-1}dx^{\mu} + {\cal L}^{\mu}_{0},
\end{eqnarray}
where ${\cal L}^{\mu}_0$ is an x-independent vector . In terms of
Moyal bracket this last equation $(4.9)$ will be rewritten as
\begin{equation}
\{x^{\mu},{\cal L}^{\nu}\}=-i{\chi}^{N}{\theta}^{{\mu}{\nu}}.
\end{equation}
Quantizing equation $(4.8)$ however will give that

\begin{equation}
[L^{\mu},\Phi]_{\times}=i{\theta}^{{\mu}{\nu}}R^N{\times}[D_{\nu},{\Phi}]_{\times}
\end{equation}
where we have used $(4.6)$ . This last equation can be iterated
to give
\begin{equation}
[D_{\mu},\Phi]_{\times}=-iR^{-N}{\theta}^{-1}_{{\mu}{\nu}}[L^{\nu},\Phi]_{\times}
+
iaR^{-N}{\theta}^{-1}_{{\mu}{\nu}}{\Delta}(R^N,R^{-N}[L^{\nu},\Phi]).
\end{equation}
This is the definition of the quantum derivations $D_{\mu}$ and
it is given in terms of the operators $L^{\mu}$ defined by
\begin{equation}
<x|L^{\mu}|x>={\cal L}^{\mu}.
\end{equation}
It must also satisfy
\begin{equation}
[X^{\mu},L^{\nu}]_{\times}=-iR^{N}{\theta}^{{\mu}{\nu}}.
\end{equation}
which follows from $(4.10)$ . Putting everything together the
Dirac operator is then given by :
\begin{equation}
D=-iR^{-N}{\theta}^{-1}_{{\mu}{\nu}}{\gamma}^{\mu}L^{\nu} +
a{\delta}D.
\end{equation}
where
\begin{equation}
[{\delta}D,\Phi]_{\times}=[F,\Phi]_{\times}+iR^{-N}{\theta}^{-1}_{{\mu}{\nu}}{\Delta}(R^N,R^{-N}[L^{\nu},\Phi]).
\end{equation}
The Dirac operator $(4.15)$ does act on the Hilbert space
\begin{equation}
H=A{\otimes}{C}^4
\end{equation}

\noindent {\bf 4.2 Renormalization And Causality} \vskip 5mm
\noindent

We define scalar fields $\hat{\Phi}$ on the quantum space-time
$QR^4$ to be elements of the algebra $A$ , they are given by
$(3.14)$ or $(3.18)$ . Action integrals for such fields will have
the form \cite{sachin}

\begin{equation}
S=\int d^4x
e^{-\frac{a{\chi}}{4}{\frac{\partial}{{\partial}x^{\mu}}}{\frac{\partial}{{\partial}x_{\mu}}}}\bigg[<x|\frac{1}{2}[D,\hat{\Phi}]_{\times}{\times}[D,\hat{\Phi}]_{\times}
- \frac{m^2}{2}{\hat{\Phi}}{\times}{\hat{\Phi}}
-\frac{g^4}{4!}({\hat{\Phi}}{\times}{\hat{\Phi}}){\times}({\hat{\Phi}}{\times}{\hat{\Phi}})|x>\bigg].
\end{equation}
The trace is taken over the coherent states basis $|x>$ .
 $m$ is the mass of the scalar field ${\hat{\Phi}}$ , $g$ is the strength of the ${\hat{\Phi}}^4$ interaction .
  These two parameters are assumed to be the physical parameters of the theory , in other words they are finite .
   The field ${\hat{\Phi}}$ is mapped via the homomorphism ${\cal X}$ to an ordinary scalar field $\hat{\phi}$ given by
\begin{equation}
\hat{\phi}(x)=e^{-\frac{a{\chi}}{4}{\frac{\partial}{{\partial}x^{\mu}}}{\frac{\partial}{{\partial}x_{\mu}}}}(<x|{\hat{\Phi}}|x>).
\end{equation}
In terms of this new field the action $(4.18)$ will read
\begin{equation}
S=\int d^4x
[\frac{1}{2}{\partial}_{\mu}{\hat{\phi}}*{\partial}^{\mu}{\hat{\phi}}
-\frac{m^2}{2}{\hat{\phi}}*{\hat{\phi}} -
\frac{g^4}{4!}({\hat{\phi}}*{\hat{\phi}})*({\hat{\phi}}*{\hat{\phi}})],
\end{equation}
The field ${\phi}$ is a general element of the algebra $({\cal
A},*)$ which is of the form $(2.22)$ . It is clearly a function of
${\chi}$ which can always be put in the form
\begin{eqnarray}
{\hat{\phi}}(x)&=&=\int \frac{d^4p}{(2{\pi})^4} {\hat{\phi}}(p,{\chi})e^{ipx}\nonumber\\
&=&{\phi} + {\hbar}{\psi}_1 +{\hbar}^2{\psi}_2 + ..+{\hbar}^N{\psi}_N.
\end{eqnarray}
${\phi}$ is a scalar field which is independent of ${\hbar}$ and
of the fuzzyness functions ${\chi}_1$ , ${\chi}_2$ ,.... ${\chi}_N$ ; in other words it is the (commutative) classical field of the theory . ${\psi}_1$ , ${\psi}_2$ ,..,
${\psi}_N$ , in the other hand , are scalar fields which are also
independent of ${\hbar}$ but do depend on 
${\chi}_1$ , ${\chi}_2$ ,..,${\chi}_N$ . Their dependence on ${\chi}'s$ is such that they go to
zero at the limit of all ${\chi}_i{\longrightarrow}0$ . ${\psi}_1$ , ${\psi}_2$ , ...${\psi}_N$ are assumed to be finite and therefore the noncommutative field ${\hat{\phi}}$ is also finite . ${\hat{\phi}}$ can then be identified with the renormalized scalar field of the theory . For simplicity we will only
consider the two-loop calculation of the ${\phi}^4$ theory . In
this case $N=2$ and we will have three scalar fields ${\phi}$ ,
${\psi}_1$ and ${\psi}_2$ and two fuzzyness functions ${\chi}_1$
and ${\chi}_2$ . The action $(4.20)$ in terms of these functions
will read

\begin{equation}
S=S[\phi]+S[\phi,{\psi}_1,{\psi}_2].
\end{equation}
For the moment we will only focus on the first term in $(4.22)$ .
The action $S[{\phi}]$ depends only on the field ${\phi}$ and it
has the form
\begin{equation}
S[\phi]=\int d^4x {\cal L} + \int d^4x {\Delta}{\cal L},
\end{equation}
where
\begin{equation}
{\cal L}=\frac{1}{2}{\partial}_{\mu}{\phi}{\partial}^{\mu}{\phi}
-\frac{m^2}{2}{\phi}^2 - \frac{g^4}{4!}{\phi}^4,
\end{equation}
and
\begin{eqnarray}
{\Delta}{\cal L}&=&\int d^4 x {\chi}{\Delta}{\cal L}_1({\phi}) + \int d^4x {\chi}^2 {\Delta}{\cal L}_2(\phi)\nonumber\\
&=&\int \frac{d^4p}{(2{\pi})^4} {\chi}(p){\Delta}{\cal
L}_1({\phi},p)\nonumber\\&+& \int \frac{d^4p}{(2{\pi})^4}
\frac{d^4l}{(2{\pi})^4} {\chi}(l){\chi}(p-l){\Delta}{\cal
L}_2({\phi},p).
\end{eqnarray}
In $(4.25)$ and in all what will follow $f(p)$ is the Fourier
transform of the function $f(x)$ and it is defined by $f(p)=\int
d^4x f(x) e^{ipx} $ . 

Although ${\phi}$ is the classical field , we will show below that ${\cal L}$ in $(4.24)$ generates exactly the renormalized action of the ${\phi}^4$ theory , and as Eq $(4.23)$ suggests ${\Delta}{\cal L}$ will generate the corresponding counter terms . ${\chi}$ will depend therefore on the classical lagrangians ${\Delta}{\cal L}_1$ ,
${\Delta}{\cal L}_2$ and the usual renormalization constants $Z_1$
, $Z_3$ and ${\delta}m^2$ . For consistency ${\Delta}{\cal L}_1$
and ${\Delta}{\cal L}_2$ should not depend on ${\chi}$ which is
the case as we can see from their explicit expressions

\begin{eqnarray}
{\Delta}{\cal L}_1(\phi,p)&=&\int \frac{d^4k}{(2{\pi})^4}{\phi}(k)\Bigg[-\frac{a}{4}(pk - k^2)(pk -k^2 + m^2){\phi}(p-k)\nonumber\\
&-&\frac{3ag^4}{4!} \int  \frac{d^4q}{(2{\pi})^4}
\frac{d^4l}{(2{\pi})^4}(ql){\phi}(p-k-q-l){\phi}(q){\phi}(l)\Bigg].
\end{eqnarray}
And
\begin{eqnarray}
{\Delta}{\cal L}_2(\phi,p)&=&\int \frac{d^4k}{(2{\pi})^4}{\phi}(k)\Bigg[\frac{1}{16}(pk-k^2+m^2)(pAk -kAk)^2{\phi}(p-k)\nonumber\\
&+&\frac{1}{4}\frac{g^4}{4!}\int\frac{d^4q}{(2{\pi})^4}\frac{d^4s}{(2{\pi})^4}\bigg[(qAs)^2+\frac{1}{2}[(p-q-s)A(q+s)]^2\bigg]\nonumber\\
&&{\phi}(p-k-q-s){\phi}(q){\phi}(s)\Bigg],
\end{eqnarray}
where $A={\theta}+ia{\eta}$ . For the ${\phi}^4$ theory it is
known that in the first order of the quantum theory both the mass
and the coupling constant need to be renormalized . In the second
order however we need also a field renormalization . So we would
have

\begin{equation}
\int \frac{d^4p}{(2{\pi})^4} {\chi}^{(1)}(p){\Delta}{\cal
L}_1({\phi},p)=\int d^4x [-\frac{1}{2}{{\delta}m}_1^2{\phi}^2
-\frac{g^4}{4!}Z_1^{(1)}{\phi}^4].
\end{equation}
and
\begin{eqnarray}
&\int& \frac{d^4p}{(2{\pi})^4} {\chi}^{(2)}(p){\Delta}{\cal
L}_1({\phi},p) + \int \frac{d^4p}{(2{\pi})^4}
\frac{d^4l}{(2{\pi})^4} {\chi}^{(1)}(l){\chi}^{(1)}(p-l){\Delta}{\cal
L}_2({\phi},p)=\nonumber\\
 &\int& d^4x[\frac{1}{2}Z_3^{(2)}{\partial}_{\mu}{\phi}{\partial}^{\mu}{\phi}-
\frac{1}{2}{{\delta}m}_2^2{\phi}^2-\frac{g^4}{4!}Z_1^{(2)}{\phi}^4].
\end{eqnarray}
${\chi}^{(1)}$ and ${\chi}^{(2)}$ are given by
${\chi}^{(1)}={\hbar}{\chi}_1$ and
${\chi}^{(2)}={\hbar}^2{\chi}_2$ . The action $(4.23)$ will then
take the form
\begin{equation}
 S[{\phi}]=\int d^4x[\frac{1}{2}Z_3{\partial}_{\mu}{\phi}{\partial}^{\mu}{\phi} -\frac{m^2+{\delta}m^2}{2}{\phi}^2 - \frac{g^4}{4!}Z_1{\phi}^4].
\end{equation}
$Z_3=1 + Z_3^{(2)}$ , $Z_1=1 +Z_1^{(1)}+Z_1^{(2)}$ and
${\delta}m^2={\delta}m_1^2 + {\delta}m_2^2$ . Solving $(4.28)$ and
$(4.29)$ for ${\chi}_1$ and ${\chi}_2$ will give

\begin{eqnarray}
{\chi}^{(1)}(p)&=&\frac{\phi(p)}{{\Delta}{\cal L}_1(\phi,p)}\Bigg[-\frac{1}{2}{\delta}m_1^2{\phi}(-p)\nonumber\\
&-& \frac{g^4}{4!}Z_1^{(1)}\int \frac{d^4k}{(2{\pi})^4}
\frac{d^4l}{(2{\pi})^4}{\phi}(k){\phi}(l){\phi}(-p-k-l)\Bigg],
\end{eqnarray}
and
\begin{eqnarray}
{\chi}^{(2)}(p)&=&\frac{\phi(p)}{{\Delta}{\cal L}_1(\phi,p)}\Bigg[\frac{1}{2}[Z_3^{(2)}p^2 - {\delta}m_2^2]{\phi}(-p)\nonumber\\
&-&\frac{g^4}{4!}Z_1^{(2)}\int \frac{d^4k}{(2{\pi})^4} \frac{d^4l}{(2{\pi})^4}{\phi}(k){\phi}(l){\phi}(-p-k-l)\Bigg]\nonumber\\
&-&\frac{{\Delta}{\cal
L}_2(\phi,p)}{{\Delta}{\cal L}_1(\phi,p)}\int
\frac{d^4l}{(2{\pi})^4}{\chi}^{(1)}(l){\chi}^{(1)}(p-l).
\end{eqnarray}
Putting the action $(4.30)$ back into $(4.22)$ we get
\begin{equation}
S=S[\hat{\phi}]+S[\hat{\phi},{\psi}_1].
\end{equation}
where $S[\hat{\phi}]$ is the action integral given by the
equation $(4.30)$ with the substitution
${\phi}{\longrightarrow}\hat{\phi}$ ,i.e
\begin{equation}
 S[\hat{\phi}]=\int d^4x[\frac{1}{2}Z_3{\partial}_{\mu}{\hat{\phi}}{\partial}^{\mu}{\hat{\phi}} -\frac{m^2+{\delta}m^2}{2}
 {\hat{\phi}}^2 - \frac{g^4}{4!}Z_1{\hat{\phi}}^4].
\end{equation}
This is exactly the standard renormalized action of the ${\phi}^4$ theory with all of its counter terms . $S[\hat{\phi},{\psi}_1]=\int d^4x {\cal L}(\hat{\phi},{\psi}_1)$
in the other hand is given by
\begin{eqnarray}
{\cal L}[\hat{\phi},{\psi}_1]&=&-\frac{a{\hbar}^2{\chi}_1}{2}{\partial}_{\mu}{\partial}_{\nu}{\hat{\phi}}
{\partial}^{\mu}{\partial}^{\nu}{\psi}_1 +\frac{a{\hbar}^2m^2{\chi}_1}{2}{\partial}_{\mu}{\hat{\phi}}
{\partial}^{\mu}{\psi}_1\nonumber\\
&+&\frac{a{\hbar}^2g^4{\chi}_1}{4}{\hat{\phi}}
^2{\partial}_{\mu}{\hat{\phi}}{\partial}^{\mu}{\psi}_1
+\frac{a{\hbar}^2g^4{\chi}_1}{4}
{\hat{\phi}}{\psi}_1{\partial}_{\mu}{\hat{\phi}}{\partial}^{\mu}{\hat{\phi}}\nonumber\\
&+&{\hbar}{\delta}m_1^2{\hat{\phi}}{\psi}_1 +
\frac{{\hbar}g^4Z_1^{(1)}}{6}{\hat{\phi}}^3{\psi}_1 .
\end{eqnarray}
As we can immediately remark , this action does not contain the field ${\psi}_2$ given in $(4.21)$ . So at this order of perturbations , ${\psi}_2$ is an arbitrary finite scalar field .

Clearly the finiteness requirement given by the equations
$(4.28)$ and $(4.29)$ reduces considerably the noncausality of the
field $\hat{\phi}$ . However as we can see from $ (4.33)$ , $(4.34)$ and $(4.35)$ this
scalar field is still highly noncausal in the sense that its conjugate momentum , which follows from the action $(4.33)$ , is not given by the ordinary expression $Z_3{\partial}_0{\hat{\phi}}$ . It does contain extra corrections coming from $(4.35)$ . Neverthless , one can construct a causal
field ${\hat{\phi}}_c$ from ${\hat{\phi}}$ as follows . The action
integral of ${\hat{\phi}}_c$ will be by definition given by

\begin{equation}
 S[{\hat{\phi}}_c]=\int d^4x[\frac{1}{2}Z_3{\partial}_{\mu}{\hat{\phi}}_c{\partial}^{\mu}{\hat{\phi}}_c -\frac{m^2+{\delta}m^2}{2}
 {\hat{\phi}}_c^2 - \frac{g^4}{4!}Z_1{\hat{\phi}}_c^4],
\end{equation}
and it should be equal to $(4.33)$ ,i.e $S[{\hat{\phi}}_c]=S$ .
The field ${\hat{\phi}}$ in the other hand will be defined by

\begin{equation}
{\hat{\phi}}_c={\hat{\phi}} + {\hbar}^2{\psi}_2^{'}.
\end{equation}
From $(4.33)$ and $(4.36)$ the field ${\psi}_2^{'}$ should satisfy
\begin{equation}
{\hbar}^2{\psi}^{'}_2=-\frac{{\cal
L}(\hat{\phi},{\psi}_1)}{({\partial}^2+m^2+\frac{g^4}{6}{\hat{\phi}}^2){\hat{\phi}}}.
\end{equation}
The field ${\hat{\phi}}_c$ is causal but not necessarily finite .
The field $\hat{\phi}$  in the other hand is finite but not causal
. Clearly the finite and causal scalar field theory which we can
construct on $QR^4$ is such that ${\hat{\phi}}_c={\hat{\phi}}$ .
The solution to this condition is clealry ${\psi}_2^{'}=0$ which
can be reexpressed as a constraint on the field ${\psi}_1$

\begin{eqnarray}
{\cal
L}(\hat{\phi},{\psi}_1)=0
\end{eqnarray}
The only consistent solution to this equation is the trivial one : $\psi_1=0$ . it is the only solution as we can check which is compatible with the field ${\hat{\phi}}$ being finite .The class of fields ${\hat{\phi}}$ given by the equations $(4.21)$
and $(4.39)$ are the only both causal and fintie scalar fields
which we can write down on $QR^4$ . The corresponding action
integral is given by the
equation $(4.20)$ or equivalently
$(4.34)$ .
\newpage
\vskip 5mm
\noindent
{\bf Remarks} \vskip 5mm
\noindent

It is very instructive to perform the following consistency check on our results . First remark that the commutation relations $(2.16)$ combined with the solutions $(4.31)$ and $(4.32)$ lead directly to the conclusion that the coordinates $x^{\mu}$ diverge , which definitely needs to be avoided in order to keep the finiteness of the field $\hat{\phi}$ . The solution to this problem is to assume that the antisymmetric tensor ${\theta}$ scales in such a way that the coordinates $x^{\mu}$ remain well defined . We write then ${\theta}=Z_{\theta}{\theta}_F$ , and compute the commutation relations $(2.16)$ which will then take the form $\{x^{\mu},x^{\nu}\}=iZ_{\theta}[a_1 Z_1^2 + a_2 {\delta}m^4 + a_3 Z_1 + a_4 {\delta}m^2 + a_5 Z_1{\delta}m^2 + a_6 Z_3 +a_7 ]{\theta}_F^{{\mu}{\nu}}$ . $a_i=a_i(x)$,$i=1,7$, are finite (computable) functions on $R^4$ . The minimal prescription that will keep this commutator from diverging is that $Z_{\theta}=1/({Z_1^2{\delta}m^4Z_3})$ . In other words ${\theta}$ measures directly the infinities of the field theory . We will leave the discussion on how really small is ${\theta}$ to a future communication .

The action $(4.18)$ or $(4.20)$ may be viewed as a classical action describing a ${\phi}^4$ theory of a (classical) noncommutative field ${\hat{\phi}}$ living on a quantum space-time $QR^4$ . The noncommutativity effects of this theory were shown to be exactly the quantum effects of an ordinary quantum ${\phi}^4$ theory of a quantum (commutative) field ${\hat{\phi}}$ living on $R^4$ . This map between the noncommutative classical field theory and the commutative quantum field theory is consistent by construction because the two limits , the classical limit ${\hbar}{\longrightarrow}0$ and the commutative limit ${\chi}{\longrightarrow}0$ , are identically the same .

\sxn{Conclusion}
\begin{itemize}

\item[-]We showed that the renormalized scalar field action on $R^4$
plus its counter terms can be rewritten only as a renormalized
action on $QR^4$ with no counter terms . This leads us to believe
that renormalization of quantum field theory is in general
equivalent to the process of quantizing the underlying space-time
.

\item[-]Finding phenomenological consequences of NCG such as the
correction to the Coulomb potential due to the noncommutativity of
space-time will be very interesting because it will allow us to
put  bounds on the nature of space-time at the very short
distances . Results will be reported elsewhere.

\item[-] Trying to include gravity as the source of the
regularization and not merely as another term in the action is
also under investigation . We would like that the commutation
relations $(2.15)$ or $(3.3)$ to be a consequence of quantum
gravity . A large extra dimension-like activity will be then used
to make quantum gravity corrections of the same order as the
quantum corrections . This will clearly involve going to higher
dimensions .

\item[-]The connection of the quantum space-time constructed in
this paper to ordinary lattices is also very important to such
matters as confinement and asymptotic freedom .
\end{itemize}

\vskip 5mm \noindent {\bf\large Acknowledgments} \vskip 5mm
\noindent I would like to thank my supervisor A.P.Balachandran for
his critical comments and remarks while this work was in progress.
This work  was supported in part by the DOE under contract number
DE-FG02-85ER40231.

\bibliographystyle{unsrt}

\end{document}